\documentclass[smallextended]{svjour3} 
\usepackage{graphicx}
\usepackage{times}
\usepackage{graphicx}
\usepackage{latexsym}
\usepackage{amsmath}
\usepackage{amssymb}

\def\lsim{\lower.5ex\hbox{$\; \buildrel < \over \sim \;$}}
\def\gsim{\lower.5ex\hbox{$\; \buildrel > \over \sim \;$}}
\begin{document}

\title{An Analytical Study on the Multi-critical Behaviour and Related 
Bifurcation Phenomena for Relativistic Black Hole Accretion}

\author{Shilpi Agarwal \and Tapas K. Das \and Rukmini Dey \and Sankhasubhra Nag}

\institute{S. Agarwal \at Faculty of Science, Banaras Hindu University, Varanasi 221005, India.\\ \email{shilpiagarwal.2006@gmail.com} \and T. K. Das \at Harish Chandra Research Institute, Allahabad-211 019, India.\\ \email{tapas@mri.ernet.in} \\ \emph{http://www.mri.ernet.in/\~\ tapas} \and R. Dey \at Harish Chandra Research Institute, Allahabad-211 019, India.\\ \email{rkmn@mri.ernet.in}\\ \emph{http://www.mri.ernet.in/\~\ rkmn} \and S. Nag \at Sarojini Naidu College for Women, Kolkata 700028, India.\\ \email{sankhasubhra\_nag@yahoo.co.in} }

\date{Received: date / Accepted: date}
\titlerunning{Multicritical Behaviour and Bifurcation in Black Hole Accretion}

\maketitle

\begin{abstract}
We apply the theory of algebraic polynomials to analytically 
study the transonic properties of general relativistic 
hydrodynamic axisymmetric accretion onto non-rotating 
astrophysical black holes. For such accretion phenomena, 
the conserved specific energy of the flow, which turns out
to be one of the two first integrals of motion in the system studied, 
can be expressed
as a 8$^{th}$ degree polynomial of the critical point of the 
flow configuration. We then construct the corresponding Sturm's 
chain algorithm to calculate the number of real roots lying
within the astrophysically relevant domain of $\mathbb{R}$.
This allows, for the first time in literature, to {\it analytically}
find out the maximum 
number of physically acceptable solution an accretion flow with 
certain geometric configuration, space-time metric, and equation of
state can have, and thus to investigate its multi-critical
properties {\it completely analytically}, 
for accretion flow in which the location of the critical points can not be 
computed without taking recourse to the numerical scheme.
This work can further be generalized to analytically calculate the 
maximal number of equilibrium points certain autonomous dynamical 
system can have in general. We also demonstrate how the transition 
from a mono-critical to multi-critical (or vice versa) flow 
configuration can be realized through the saddle-centre 
bifurcation phenomena using certain techniques of the catastrophe theory.

\end{abstract}


\maketitle

\section{Introduction}
\label{section1}
\noindent
In order to satisfy the inner boundary conditions imposed by the event horizon,
accretion onto astrophysical black holes exhibit transonic properties in 
general \cite{lt80}. A physical transonic accretion solution can mathematically 
be realized as critical solution on the phase portrait 
\cite{rb02,ap03,ray03a,ray03b,rbcqg05a,rbcqg05b,crd06,rbcqg06,rbcqg07a,br07,gkrd07,jkb09}. 
Multi-critical accretion may be referred to the specific 
category of accretion flow configuration having multiple critical points 
accessible to the accretion solution. For certain astrophysically relevant 
values of the initial boundary conditions, low angular momentum sub-Keplerian 
axisymmetric black hole accretion can have at most three critical points 
all together -- where two saddle type critical points accommodate one centre 
type critical point in between them
\cite{lt80,az81,boz-pac,boz1,fuk83,fuk87,fuk04,fuk04a,lu85,lu86,bmc86,ak89,abram-chak,ky94,yk95,caditz-tsuruta,das02,bdw04,abd06,dbd07,das-czerny,swagata}. 
Transonic solution passing through the aforementioned two critical points 
can be joined through a stationary shock generated as a consequence 
of the presence of the angular momentum barrier
\cite{fuk87,c89,das02,dpm03,dbd07,das-czerny}. The existence of 
such weakly rotating accretion in realistic astrophysical 
environment have also been observed \cite{ila-shu,liang-nolan,bisikalo,ila,ho,igu}.
A complete investigation of the multi-critical shocked accretion 
flow around astrophysical black holes necessitates the numerical integration 
of the nonlinear stationary equations describing the velocity phase 
space behaviour of the flow. 

However, for all the importance of transonic flows, there exists as yet no 
general mathematical prescription allowing one a direct analytical understanding 
of the nature of the multi-criticality without having to take recourse to the existing 
semi-analytic approach of numerically finding out the total number of 
physically acceptable critical points the accretion flow can have. 

This is precisely the main achievement of our work presented in this paper. Using the
theory of algebraic polynomials, we developed a mathematical algorithm capable of
finding the number of physically acceptable solution a polynomial  can have, for
any arbitrary large value of $n$ ($n$ being the degree of the polynomial).
For a specified set of values of the initial boundary conditions, 
we mathematically predict whether the flow will be multi-critical (more than
one real physical roots for the polynomial) or not. 
This paper, thus, purports to address that particular issue of investigating the transonicity 
of a general relativistic flow structure around non rotating black holes
without encountering the usual semi-analytic numerical
techniques, and to derive some predictive insights about the qualitative character of the 
flow, and in relation to that, certain physical features of the multi-criticality of 
the flow will also be addressed. In our work, we would like to develop a complete
analytical formalism to investigate the critical behaviour
of the general relativistic low angular momentum inviscid axisymmetric advective hydrodynamic accretion flow around a
non rotating black hole. 

To accomplish the aforementioned task, we 
first construct the equation describing the space gradient
of the dynamical flow velocity of accreting matter. Such equation
is isomorphic to a first order autonomous dynamical system.
Application of the fixed point analysis enables to construct
an $8$th degree algebraic equation for the space variable
along which the flow streamlines are defined to
possess certain first integrals of motion. The constant coefficients for
each term in that equation are functions of astrophysically 
relevant initial boundary conditions. Such initial
boundary conditions span over a certain domain on the real
line $\mathbb{R}$ -- effectively, as individual sub-domain of 
${\mathbb{R}}{\times}\mathbb{R}{\times}\mathbb{R}$ 
for the polytropic accretion.
The solution of aforesaid equation would then provide the critical
(and consequently, the sonic) point $r_c$. The critical points itself
are permissible only within a certain open interval
$\left]r_g,L_{{\rightarrow}{\infty}}\right[$, 
where $r_g$ is the
radius of the event horizon and $L_{{\rightarrow}{\infty}}$ is the
physically acceptable maximally allowed limit on the value of 
a critical point. 

Since for polynomials of degree $n>4$, analytical solutions are 
not available, we use the Sturm's theorem (a corollary of 
the Sylvester's theorem), to construct the Sturm's chain algorithm, which
can be used to calculate the number of real roots (lying within a
certain sub-domain of $\mathbb{R}$) for a polynomial of any countably 
finite arbitrarily large integral $n$, subjected to certain sub-domains
of constant co-efficients. The problem now reduces to identify the polynomials 
in $r_c$ with the Sturm's sequence, and to find out the maximum 
number of physically acceptable solution an accretion flow with 
certain geometric configuration, space-time metric, and equation of
state can have, and thus to investigate its multi-critical
properties completely analytically, 
for accretion flow in which the critical points can not be computed analytically.
Our work, as we believe, has significant importance, because for the 
first time in the literature, we provide a purely analytical method,
by applying certain theorem of algebraic polynomials
to check whether certain astrophysical hydrodynamic 
accretion may undergo more than one sonic transitions.

We further demonstrate how the transition of number of critical points may be taken into account 
considering the bifurcation phenomenon in the parameter space. The transition of number of 
critical points in this case is associated with the merging and destruction (or emergence 
and separating apart, viewing in the other way round) of a saddle-centre pair, i.e. a saddle-centre 
bifurcation common in conservative systems, which may be tracked down using technique of 
catastrophe theory. The bifurcation lines in the parameter space exactly conform with  
the transition boundaries of the across which the number of critical points changes.

\section{First Integral of Motion as a Polynomial in Critical Radius}
\label{section2}
\noindent
Following standard literature, we assume that the axisymmetric accretion flow has a radius dependent local thickness $H(r)$, 
and its central plane coincides with the equatorial plane of the black hole. It is common practice in accretion disc theory (\cite{mkfo84,pac87,acls88,ct93,ky94,abn96,nkh97,wiita99,hawley-krolik,armitage}) to use the vertically integrated model in describing the black hole accretion discs where the equations of motion apply to the equatorial plane of the black hole assuming the flow to be in hydrostatic equilibrium along transverse direction. We follow the same procedure here. The thermodynamic flow variables are 
averaged over the disc height, i.e. a thermodynamic quantity $y$ used in our model is vertically integrated over the disc height and averaged as $\bar{y}=\int^{H(r)}_0y dh/\int^{H(r)}_0dh$. 

We follow \cite{alp97} to derive an expression for the disc height $H(r)$ in our geometry since the relevant equations in \cite{alp97} are non-singular on the horizon and can accommodate both axial and quasi spherical flow geometry. The disc height comes out to be \cite{dbd07},
\begin{equation}
H(r)=\frac{c_sr}{\lambda}\sqrt{\frac{2(\gamma-1)(1-u^2)[r^3-\lambda^2(r-2)]}{\gamma[\gamma-(1+c_s^2)](r-2)}}
\end{equation}
where $\lambda$ and $\gamma$ are the specific flow angular momentum and the adiabatic index
of the flow, respectively.  $u$ and $c_s$ being the dynamical flow velocity and the speed of 
propagation of the acoustic perturbation (adiabatic sound speed) embedded within the accretion flow.
In this work, we employ polytropic accretion. However, polytropic accretion is not the only choice to describe the general relativistic axisymmetric  black-hole accretion. Equations of state other than the adiabatic one, such as the isothermal equation \cite{yk95} or two temperature plasma \cite{man2000} have also been used to study the black-hole accretion flow. 

\noindent
For accretion flow of aforementioned category, two first integrals of motion along the 
streamline, viz, the dimensionless conserved specific flow energy i.e., the energy per unit 
mass which actually is scaled by the rest mass of the flow  ${\cal E}$, and the mass accretion rate
${\dot M}$, may be obtained as (the radial distance $r$ here is actually scaled by the 
factor $GM_{BH}/c^2$, and all the velocities, both $u$ as well as $c_s$ have been
scaled by the velocity of light $c$ in vacuum. $M_{BH}$ is the mass of the black hole. Natural geometric 
unit has been used where the values of all fundamental constants have been taken to be unity,
see, e.g., \cite{dbd07} for further detail)
\begin{equation}
{\cal E}=\left[ \frac{(\gamma -1)}{\gamma -(1+c^{2}_{s})}
 \right]r
\sqrt{\frac{r-2}{r^3-\lambda^2\left(r-2\right)}}
\frac{1}{\sqrt{1-u^2}},
\label{eq1}
\end{equation}
\begin{equation}
{\dot M}= 
\frac{4{\pi}{\rho}c_sr^{\frac{3}{2}}u}{\lambda}
\sqrt{
\frac{2\left(\gamma-1\right)\left[r^3-\lambda^2\left(r-2\right)\right]}{\gamma\left[\gamma-\left(1+c_s^2\right)\right]}
}\, ,
\label{eq2}
\end{equation}
where $\rho$ is the mass density. The expression for ${\cal E}$ is obtained by 
integrating the stationary part of the Euler equation and the expression for 
${\dot M}$ is obtained by integrating the stationary part of the 
continuity equation (by properly taking care of the flow thickness). 
The conserved specific entropy accretion rate ${\dot {\cal M}}$ is computed as 
a quasi constant multiple of ${\dot M}$ as:
\begin{equation}
{\dot {\cal M}}=
 4\pi \left( \frac{1}{\lambda} \sqrt\frac{2}{\gamma} \right)
\left[\frac{c_{s}} {\left (1-\frac{c_{s}^2}{\gamma-1}\right )^{\frac{1}{2}}}
 \right]^{\frac{\gamma+1}{\gamma-1}} u r \left[r^4-\lambda^2 r(r-2)
 \right]^\frac{1}{2},
\label{eq3}
\end{equation}
\noindent
We thus have two primary first integrals of motion along the
streamline -- the specific energy of the flow ${\cal E}$ and the
mass accretion rate ${\dot M}$. Even in the absence of creation
or annihilation of matter, the entropy accretion rate ${\dot {\cal M}}$
is not a generic first integral of motion. As the expression for
${\dot {\cal M}}$
contains the quantity $K{\equiv}p/{\rho}^{\gamma}$ ({$p$ being the flow 
pressure}), which is a measure
of the specific entropy of the flow, the entropy accretion rate ${\dot {\cal M}}$
remains constant throughout the flow only if the entropy per particle remains
locally invariant. This condition may be violated if the accretion is
accompanied by a shock. Thus ${\dot {\cal M}}$
is conserved for shock free polytropic accretion and
becomes discontinuous (actually, increases) at the shock location,
if such a shock is formed.

\noindent
The gradient of the acoustic velocity $c_s$ as well as the dynamical velocity $u$ can 
be obtained by differentiating the expression for the entropy accretion rate 
and the mass accretion rate respectively:
\begin{equation}
\frac{dc_{s}}{dr}=-\frac{c_{s}(\gamma-1)
\left[\gamma-(1+c^{2}_{s}\right)]}{(\gamma+1)}
\left[ \frac{1}{u} \frac{du}{dr} + {f_{1}}(r,\lambda) \right],
\label{eq3a}
\end{equation}
where
\begin{equation}
{f_{1}}(r,\lambda) = \frac{3r^{3}-2\lambda^{2}r+
3{\lambda^2}}{r^{4}-\lambda^{2}r(r-2)}\, .
\label{eq4}
\end{equation}
\begin{equation}
\frac{du}{dr} = \frac{(\frac{2}{\gamma+1})c^{2}_{s} {f_{1} } (r,\lambda)
 - {f_{2} } (r,\lambda)}{\frac{u}{1-u^{2}} - \frac{2c^{2}_{s}}{u(\gamma+1)}}
=\frac{{\cal N}\left(r,\lambda,c_s\right)}{{\cal D}\left(u,c_s\right)}\, ,
\label{eq5}
\end{equation}
where
\begin{equation}
{f_{2} } (r,\lambda) = \frac{2r-3}{r(r-2)} -
\frac{2r^{3}-\lambda^{2}r+\lambda^{2}}{r^{4}-\lambda^{2}r(r-2)}\, .
\label{eq6}
\end{equation}
A real physical transonic flow must be smooth everywhere,
except possibly at a shock. Hence, if the denominator
${{\cal D}\left(u,c_s\right)}$ of Eq. (\ref{eq5})
vanishes at a point, the numerator
${{\cal N}\left(r,\lambda,c_s\right)}$ must also vanish at that point to ensure the
physical continuity of the flow. One therefore arrives at the {\em critical point}
conditions
by  making ${{\cal D}\left(u,c_s\right)}$  and
${{\cal N}\left(r,\lambda,c_s\right)}$ of Eq. (\ref{eq5})
simultaneously equal to zero.
 We thus obtain the
critical point conditions as
\begin{equation}
u_c=\pm \sqrt{
\frac
{{f_{2}}(r_c,\lambda)}
{{{f_{1}}(r_c,\lambda)}+{{f_{2}}(r_c,\lambda)}}
}; \;\;\;\;
c_c=\pm\sqrt{\frac{\gamma+1}{2}\left[\frac
{{f_{2}}(r_c,\lambda)}{{f_{1}}(r_c,\lambda)}
\right]};
\label{eq41}
\end{equation}
where $u_c\equiv u({r_c})$ and $c_c\equiv c_s (r_c)$, $r_c$ being the
 location
of the critical point. $f_1(r_c,\lambda)$ and $f_2(r_c,\lambda)$
are defined as:
\begin{equation}
{f_{1}}(r_c,\lambda) = \frac{3r_c^{3}-2\lambda^{2}r_c+
3{\lambda^2}}{r_c^{4}-\lambda^{2}r_c(r_c-2)},~
{f_{2} } (r_c,\lambda) = \frac{2r_c-3}{r_c(r_c-2)} -
\frac{2r_c^{3}-\lambda^{2}r_c+\lambda^{2}}{r_c^{4}-\lambda^{2}r_c(r_c-2)}
\label{eq42}
\end{equation}
Clearly, the critical points are not coincident with the sonic points since $M_c=\left({u_c}/{c_c}\right)< 1$. 
This is a consequence of the choice of the equation of state. The adiabatic equation of 
state used in this work produces non constant (with respect to the radial space direction) sound speed. Since 
the disc height contains the sound speed and the thermodynamic quantities calculated in the 
accretion flow have been averaged over the flow thickness, non constant sound speed accounts for the 
non-isomorphism of the critical points and the sonic points. If one uses the sound 
speed obtained from isothermal equation of state, or a flow geometry different from the configuration in the vertical equilibrium as has been 
assumed here, the critical points will coincide with the sonic points, see, e. g., \cite{abd06,swagata} 
for further detail.

We substitute the explicit value of $u_c$ and $c_c$ from Eq. (\ref{eq41}) to the expression for 
the specific energy ${\cal E}$ in Eq. (\ref{eq1}) to
derive the explicit form of the energy first integral polynomial in $r_c$ as:
\begin{eqnarray}
 &&r_c^{8} \{-36\left( -1 + {\cal E}^2 \right) {\left( -1 + \gamma  \right) }^2 \}
+ r_c^{7} \{   12\left( -1 + \gamma  \right) 
   \left( -17\left( -1 + \gamma  \right)  + 
     {\cal E}^2\left( -11 + 13\gamma  \right)  \right) \} +
\nonumber  \\   &&  
r_c^{6} \{  -24{\left( -1 + \gamma  \right) }^2\left( -16 + {\lambda }^2 \right)  + 
   {\cal E}^2\left( -121 + 60{\lambda }^2 + 
      \gamma \left( 286 - 96{\lambda }^2 \right)  + 
      {\gamma }^2\left( -169 + 36{\lambda }^2 \right)  \right) \} 
\nonumber \\  &&  
+ r_c^{5} \{ -2\left( 120 + \left( -86 + 163{\cal E}^2 \right) {\lambda }^2 + 
     {\gamma }^2\left( 120 + \left( -86 + 99{\cal E}^2 \right) {\lambda }^2
        \right)  - 2\gamma \left( 120 + 
        \left( -86 + 133{\cal E}^2 \right) {\lambda }^2 \right)  \right) \} 
\nonumber \\ && \;\;
+ r_c^{4}\{  {\lambda }^2\left( -460{\left( -1 + \gamma  \right) }^2 + 
     {\cal E}^2\left( 588 - 25{\lambda }^2 + 
        {\gamma }^2\left( 356 - 9{\lambda }^2 \right)  + 
        \gamma \left( -976 + 30{\lambda }^2 \right)  \right)  \right) \}
\nonumber  \\  &&\;\;\; \;
+ r_c^3  \{ 4{\lambda }^2\left( 136{\left( -1 + \gamma  \right) }^2 + 
     {\cal E}^2\left( -88 + 45{\lambda }^2 + 
        \gamma \left( 148 - 52{\lambda }^2 \right)  + 
        {\gamma }^2\left( -52 + 15{\lambda }^2 \right)  \right)  \right)  \} 
\nonumber \\ && \;\;\; \;
+ r_c^2  \{ -4{\lambda }^2\left( 60 + 121{\cal E}^2{\lambda }^2 + 
     {\gamma }^2\left( 60 + 37{\cal E}^2{\lambda }^2 \right)  - 
     2\gamma \left( 60 + 67{\cal E}^2{\lambda }^2 \right)  \right) \} 
\nonumber \\ && \;\;\; \;\; 
+ r_c  \{32{\cal E}^2\left( 18 - 19\gamma  + 5{\gamma }^2 \right) {\lambda }^4\} 
+\{ -64{\cal E}^2{\left( -2 + \gamma  \right) }^2{\lambda }^4\}  
=0
\label{eq43}
\end{eqnarray}
The above equation, being an $n=8$ polynomial, is non analytically solvable. Being equipped with the 
details of the Sturm theorem and its appropriate application in the next section (\S \ref{section3}),
in 
\S \ref{section5}
we will demonstrate how we can analytically find out the number of physically admissible 
real roots for this polynomial, and can investigate the transonicity of the flow. 

\section{Sturm theorem and generalized sturm sequence (chain)}
\label{section3}
\noindent
In this section we will elaborate the idea of the generalized Strum sequence/chain, and will discuss
its application in finding the number of roots of a algebraic polynomial equations with real co-efficients.
Since the central concept of this  theorem is heavily based on the idea of the greatest common 
divisor of a polynomial and related Euclidean algorithm, we start our discussion by clarifying 
such concept in somewhat great detail for the convenience of the reader.
\subsection{Greatest common divisor for two numbers}
\label{subsection3.1}
\noindent

Given two non-zero integers $z_1$ and $z_2$, one defines that $z_1$ divides $z_2$, if and 
only if there exists some integer $z_3{\in{\mathbb Z}}$ such that:
\begin{equation}
z_2=z_3z_1
\label{st1}
\end{equation}
The standard notation for the divisibility is as follows:
\begin{gather}
z_1{{\vert}}z_2 \text{~means~`} z_1 \text{~divides~} z_2  \text{'}
\label{st2}
\end{gather}
The concept of divisibility applies to the polynomials as well, we treat such situations
in the subsequent paragraphs. 

Now consider two given integers $z_1$ and $z_2$, with at least one of them being a non-zero 
number. The `greatest common divisor' (or the `greatest common factor' or the 
`highest common factor') of $z_1$ and $z_2$, denoted by $gcd(z_1,z_2)$, is the positive 
integer $z_d{\in}{\mathbb Z}$, which satisfies:
\begin{eqnarray}
{\rm i)} z_d{\vert}z_1 ~{\rm and}~ z_d{\vert}z_2.
\nonumber \\
{\rm ii)} {\rm For~any~other}~z_c{\in}{\mathbb Z},~if~z_c{\vert}z_1~{\rm and} z_c{\vert}z_2
\nonumber \\
{\rm then} z_c{\vert}z_d
\label{st3}
\end{eqnarray}
In other words, the greatest common divisor $gcd(z_1,z_2)$ of two non zero integers $z_1$ 
and $z_2$ is the largest possible integer that divides both the integers without 
leaving any remainder. Two numbers $z_1$ and $z_2$ are called `co-prime] (alternatively, `relatively prime'),
if:
\begin{equation}
gcd(z_1,z_2)=1
\label{st4}
\end{equation}
The idea of a greatest common divisor can be generalized by defining the greater 
common divisor of a non empty set of integers. If ${\cal S_{\mathbb Z}}$ is a non-empty set of 
integers, then the greatest common divisor of ${\cal S_{\mathbb Z}}$ is a positive integer
$z_d$ such that:
\begin{eqnarray}
{\rm i)~If}~z_d{\vert}z_1{\rm for~all}~z_1{\in}{\cal S_{\mathbb Z}}
\nonumber \\
{\rm ii) If}~z_2{\vert}z_1,~{\rm for~all}~z_1{\in} {\cal S_{\mathbb Z}},~{\rm then}~z_2{\vert}z_d
\label{st4a}
\end{eqnarray}
then we denote $z_d=gcd({\cal S_{\mathbb Z}})$.
\subsection{Euclidean algorithm}
\label{subsection3.2}
\noindent
Euclidean algorithm (first described in detail in Euclid's `Elements' in 300 BC, and is still 
in use, making it the oldest available numerical algorithm still in common use) provides an 
efficient procedure for computing the greatest common divisor of two integers. Following Stark
{\cite{stark}}, below we provide a simplified illustration of the Euclidean algorithm for 
two integers:

Let us first set a `counter' $i$ for counting the steps of the algorithm, with initial 
step corresponding to $i=0$. Let any $i$th step of the algorithm begins with two non-negative
remainders $r_{i-1}$ and $r_{i-2}$ with the requirement that
$r_{i-1}<r_{i-2}$, owing to the fact that the fundamental aim of the algorithm is to 
reduce the remainder in successive steps, to finally bring it down to the zero in the 
ultimate step which terminates the algorithm. Hence, for the dummy index $i$, at the first step we have:
\begin{equation} 
r_{-2}=z_2~{\rm and}~r_{-1}=z_1
\label{st5}
\end{equation}
the integers for which the greatest common divisor is sought for. After we divide $z_2$ by 
$z_1$ (operation corresponds to $i=1$), since $z_2$ is not divisible by $z_1$, one obtains:
\begin{equation}
r_{-2} = q_0 r_{-1} + r_{0}
\label{st6}
\end{equation}
where $r_0$ is the remainder and $q_0$ be the quotient. 

For any arbitrary $i$th step of the algorithm, the aim is to find a quotient $q_j$ and remainder
$r_i$, such that:
\begin{equation}
r_{i-2}=q_ir_{i-1}+r_i,~{\rm where~r_i<r_{i-1}}
\label{st7}
\end{equation}
at some step $i=j$ (common sense dictates that $j$ can not be infinitely large), the 
algorithm terminates because the remainder becomes zero. Hence the final 
{\it non-zero} remainder $r_{j-1}$ will be the greatest common divisor of the 
corresponding integers. 

We will now illustrate the Euclidean algorithm for finding the greatest common divisor for two 
polynomials.
\subsection{Greatest common divisor and related Euclidean algorithm for polynomials}
\label{subsection3.3}
\noindent
Let us first define a polynomial to be `monic' if the co-efficient of the term 
for the highest degree variable in the polynomial is unity (one). Let us now consider $p_1(x)$ and $p_2(x)$
to be two nonzero polynomials with co-efficient from a field ${\mathbb F}$ (field of real, complex,
or rational numbers, for example). A greatest common divisor of $p_1(x)$ and $p_2(x)$ is defined to 
the the monic polynomial $p_d(x)$ of highest degree such that $p_d(x)$ divides both $p_1(x)$
and $p_2(x)$. It is obvious that ${\mathbb F}$ be field and $p_d(x)$ be a monic, are 
necessary hypothesis. 

In more compact form, a greatest common divisor of two polynomials $p_1,p_2{\in}{\mathbb R}[{\mathbb X}]$
is a polynomial $p_d{\in}{\mathbb R}[{\mathbb X}]$ of greatest possible degree which divides 
both $p_1$ and $p_2$. Clearly, $p_d$ is not unique, and is only defined upto multiplication 
by a non zero scalar, since for a non zero scalar $c{\in}{\mathbb R}$, if $p_d$ is a 
$gcd(p_1,p_2{\in}{\mathbb R}[{\mathbb X}]$), so as $cp_d$.  Given polynomials 
$p_1,p_2{\in}{\mathbb R}[{\mathbb X}]$, the division algorithm provides polynomials 
$p_3,p_4{\in}{\mathbb R}[{\mathbb X}]$, with $deg(p_4)<deg(p_3)$ such that 
\begin{equation}
p_1=p_3p_2+p_4
\label{st8}
\end{equation}
Then, if $p_d$ is $gcd(p_1,p_2)$, if and only if $p_d$ is $gcd(p_2,p_4)$ as is obvious.

One can compute the $gcd$ of two polynomials by collecting the common factors by factorizing 
the polynomials. However, this technique, although intuitively simple, almost always 
create a serious practical threat while making attempt to factorize the large high degree 
polynomials in reality. Euclidean algorithm appears to be relatively less complicated and 
a faster method for all practical purposes. Just like the integers as shown in the previous 
subsection, Euclid's algorithm can directly be applied for the polynomials as well, with decreasing 
degree for the polynomials at each step. The last non-zero remainder, after made monic if necessary, comes out 
to be the greatest common divisor of the two polynomials under consideration.

Being equipped with the concept of the divisibility, $gcd$ and the Euclidean algorithm, we 
are now in a position to define the Strum theorem and to discuss its applications. 
\subsection{The Sturm Theorem: The purpose and the definition}
\label{subsection3.4}
\noindent
The Sturm theorem is due to Jacaues Charles Francois Sturm, 
a Geneva born French mathematician and a close collaborator of Joseph Liouville. 
The Sturm theorem, published in 
1829 in the eleventh volume of the `Bulletin des Sciences de Ferussac' under the
title `Memoire sur la resolution des equations numeriques' \footnote{According to some 
historian, the theorem was originally discovered by Jean Baptist Fourier, well before 
Sturm, on the eve of the French revolution.}. The Sturm theorem, which is actually a root 
counting theorem, is used to find the number of real roots over a certain interval 
of a algebraic polynomial with real co-efficient. It can be stated as:
\begin{theorem}
The number of real roots of an algebraic polynomial with real 
coefficient whose roots are simple over an interval, the endpoints of which 
are not roots, is equal to the difference between the number of sign changes
of the Sturm chains formed for the interval ends.
\end{theorem}
Hence, given a polynomial $p{\in}{\mathbb R}[{\mathbb X}]$, if we need to find 
the number of roots it can have in a certain open interval $]a,b[$, $a$ and $b$ 
not being the roots of $f$, we then construct a sequence, called `Sturm chain', of
polynomials, called the generalized strum chains. Such a sequence is derived from 
$p$ using the Euclidean algorithm. For the polynomial $p$ as described above,
the Sturm chain $p_0,p_1...$ can be defined as:
\begin{eqnarray}
p_0 = p
\nonumber \\
p_1 = p^{\prime}
\nonumber \\
p_n=-{\rm rem}\left(p_{n-2},p_{n-1}\right), n{\ge}2
\label{st8a}
\end{eqnarray}
where $rem\left(p_{n-2},p_{n-1}\right)$ is the remainder of the polynomial 
$p_{n-2}$ upon division by the polynomial $p_{n-1}$. 
The sequence terminates once one of the $p_i$ becomes zero. We then evaluate this chain 
of polynomials at the end points $a$ and $b$ of the open interval. The number of roots of $p$ 
in $]a,b[$ is the difference between the number of sign changes on the chain of 
polynomials at the end point $a$ and the number of sign changes at the end 
point $b$. Thus, for any number $t$, if $N_{p(t)}$ denotes the number of sign changes in the 
Sturm chain $p_0(t),p_1(t),...$, then for real numbers $a$ and $b$ that (both) are not 
roots of $p$, the number of distinct real roots of $p$ in the open interval 
$]a,b[$ is $\left[N_{p(a)}-N_{p(b)}\right]$. By making $a{\rightarrow}{-\infty}$
and $b{\rightarrow}{+\infty}$, one can find the total number of roots $p$ can have on the 
entire domain of ${\mathbb R}$. 

A more formal definition of the Strum theorem, as a corollary of the Sylvester's theorem,
is what follows:

{\bf Definition } Let $R$ be the real closed field, and let $p$ and
$P$ be in $R[X]$.The Sturm sequence of $p$ and $P$ is the sequence
of polynomials $({p_0},{p_1},...,{p_k})$ defined as follows:

${p_0}=p$, ${p_1}=p'P$

${p_i}={p_{i-1}}{q_i}-{p_{i-2}}$ with ${q_i}\in R[X]$ and
deg$({p_i})<deg({p_{i-1}})$ for $i=2,3,...,k$, ${p_k}$ is a
greatest common divisor of $p$ and $p'P$.

 Given a sequence
$({a_0},...,{a_k})$ of elements of $R$ with ${a_0} \neq 0$, we
define the number of sign changes in the sequence
$({a_0},...{a_k})$ as follows: count one sign change if
${a_i}{a_l}<0$ with $l=i+1$ or $l>i+1$ and ${a_j}=0$ for every $j$, $i<j<l$.

If $a \in R$ is not a root of $p$ and $({p_0},...,{p _k})$
is the Sturm sequence of $p$ and $P$, we define $v(p,P;a)$ to be the
number of sign changes in $({p_0}(a),...{p_k}(a))$.

\begin{theorem}  {\bf (Sylvester's Theorem\footnote{As stated 
in \cite{RAG}.})}
Let $R$ be a real closed field and let $p$ and $P$ be two polynomials in
$R[X]$. Let $a,b \in R$ be such that $a < b$ and neither $a$ nor $b$ are
roots of $p$. Then the difference between the number of roots of $p$ in
the interval $]a,b[$ for which $P$ is positive and the number of roots of
$p$ in the interval $]a,b[$ for which $P$ is negative, is equal to
$v(p,P;a) - v(p,P;b)$

\end{theorem}

\begin{corollary} {\bf (Sturm's Theorem):}
Let $R$ be a real closed field and $p\in R[X]$. Let $a,b\in R$ be
such that $a<b$ and neither $a$ nor $b$ are roots of $p$. Then the
number of roots of $p$ in the interval $]a,b[$ is equal to
$v(p,1;a)-v(p,1;b)$.
\end{corollary}

The proof of these two theorems are given in the Appendix I.

\section{Number of available critical points for relativistic accretion}
\label{section5}
\noindent
We first write down the
complete expression for the Sturm chains. Then for a suitable parameter set 
$\left[{\cal E},\lambda,\gamma\right]$, we can find the difference of the sign change of 
the Sturm chains at the open interval left boundary, i.e., at the event horizon 
and at the right boundary, i.e., at some suitably chosen large distance, say,
$10^8$ gravitational radius (which is such a large 
distance that beyond which practically no critical point is expected to form unless the 
specific flow energy has an extremely low value, i.e., very cold accretion flow), 
to find the number of 
critical points the accretion flow can have. 

The form of the original polynomial has already been explicitly expressed using left hand side of
Eq.~\ref{eq43}. We now construct the Sturm chains as:
\begin{eqnarray*}
p_0(r) &=& a_8 r^8 + a_7 r^7 + a_6 r^6 + a_5 r^5 + a_4 r^4 + a_3 r^3 + a_2 r^2 + a_1 r + a_0 
\nonumber \\
p_1(r) &=& 8 a_8 r^7 + 7a_7 r^6 + 6 a_6 r^5 + 5 a_5 r^4 + 4 a_3 r^3 + 3 a_3 r^2 + 2 a_2 r + a_1 
\nonumber \\
p_2(r) &=& - rem(p_0/p_1)= c_6 r^6 + c_5 r^5 + c_4 r^4 + c_3 r^3 + c_2 r^2 + c_1 r + c_0\\ &&\textrm{(the negative of the remainder of division of 
$p_0$ by $p_1$)}
\nonumber \\
p_3(r) &=& -rem (p_1/p_2) = d_5 r^5 + d_4 r^4 + d_3 r^3 + d_2 r^2 + d_1 r + d_0 \nonumber\\
p_4(r) &=& -rem(p_2/p_3) = e_4 r^4 + e_3 r^3 + e_2 r^2 + e_1 r + e_0 \nonumber\\
p_5(r) &=& -rem(p_3/p_4) = f_3 r^3 + f_2 r^2 + f_1 r + f_0\nonumber\\
p_6(r) &=& -rem(p_4/p_5) =  g_2 r^2 + g_1 r + g_0 \nonumber\\
p_7(r) &=& -rem(p_5/p_6) = h_1 r + h_0\nonumber\\
p_8 (r) &=& - rem(p_6/p_7) = i_0\nonumber\\
\label{st15}
\end{eqnarray*}
Where the explicit expression of the corresponding co-efficients $a_i, c_i, d_i ...$ has been provided 
in the equation (\ref{eq43}) and in the Appendix - II.  If one needs to figure out the number of roots of $p_0$ in $[a,b]$, 
the number of sign changes in the sequence  
$p_0(a), p_1(a), p_2(a), p_3(a), p_4(a), p_5(a), p_6(a), p_7(a), p_8(a)$ is to be counted and let us call it $v(p_0, a)$. Similarly,  
the count the number of sign changes in the sequence  $p_0(b), p_1(b), p_2(b), 
p_3(b), p_4(b), p_5(b), p_6(b), p_7(b), p_8(b)$  is to be called as $v(p_0, b)$.
Then, the number of roots of $p_0$ in $[a,b]$ is  $v(p_0, a) - v(p_0, b).$

It is important to note that direct application of 
the Sturm's theorem may not always be sufficient since some of the roots may 
yield a negative energy for ${\cal E}$ (since the ${\cal E}$ equation was 
squared to get the polynomial). Since we are interested in accretion with the 
positive positive Bernoulli's constant, to get positive values of the energy, 
we must impose the condition that
\begin{equation}
\gamma
-(1+ c_{s}^{2}) \geq 0, 
\label{st16}
\end{equation}
which is the term present in ${\cal E}$ which could
go negative.
This introduces the condition that $\displaystyle\frac{p(r)}{q(r)} \geq 0$,  where
$p(r)$ and $q(r)$ are $4$th order polynomials given by,
\begin{subequations}
\begin{eqnarray}
  p(r) &=& 6(\gamma-1)r^4-(11\gamma-13)r^3\nonumber \\&&-(5\gamma-3)\lambda^2r^2+2(9\gamma-5)\lambda^2r-8(2\gamma-1)\lambda^2,\\
  q(r) &=& 6r^4-12r^3-4\lambda^2r^2+14\lambda^2r-12\lambda^2.
\end{eqnarray}\label{st17}
\end{subequations}
To find the region where this happens, 
one has to find the $4$ roots of each of $p(r)$ and $q(r)$ -- which is analytically 
possible since roots of quartics are analytically solvable. Once the roots are obtained 
it is a trivial matter to check for what regions the rational function is positive.

\begin{figure}[htb]
	\centering
		\includegraphics[width=0.65\textwidth]{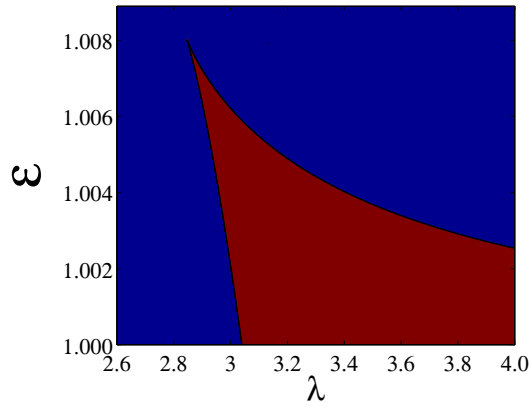}
	\caption{The lighter region (online version red) corresponds to 3 roots and the darker shade (online version blue) indicates 1 root only. The value of $\gamma$ is $4/3$. }
	\label{fig:sturmf}
\end{figure}
A simplified version for the above mentioned procedure to find the positivity 
condition is as follows:

We would like to find out the intervals in which $p(r)/q(r) >0$ where $p(r)$ and
$q(r)$ are quartic polynomials. We factorize $p(r) =
(r-r_1)(r-r_2)(r-r_3)(r-r_3)$ and $q(r)= (r-s_1)(r-s_2)(r-s_3)(r-s_4)$
using the algorithm for finding roots of a quartic. If the roots are all
real, we note down the sign changes of each factor to the right
and left of each root and find out the intervals where the rational
function is positive. If there are complex roots, they come in complex
conjugates, since the coefficients of the polynomials are real. Say, if
$r_3$ is complex and $r_4$ is its complex conjugate, then the part
$(r-r_3)(r-r_4) = r^2 - (r_3+r_4)r + r_3 r_4$ does not change sign since it is
non-zero on the real line. It is easy to determine its sign.

To demonstrate the procedure described above,  the number of roots of the $8$th order polynomial $p_0$ (in the Strum sequence) within 
the admissible range of $\cal{E}, \lambda$ and $\gamma$ (usually 
by keeping the value of $\gamma$ to be fixed to obtain a two dimensional 
parameter space)
are evaluated explicitly and that shows two distinct regions in $\cal{E} -\lambda$ space (see Fig.~\ref{fig:sturmf}).
The wedge shaped region corresponds to 3 roots implying 3 critical points and the rest of the  parametric space corresponds to 
single root implying only one critical point. This feature emerging from the above mentioned algorithm, exactly conforms with
the numerical results (using the explicit root finding methods) available in the current literature \cite{dbd07}. It may be worthwhile to mention here that in addition to these  roots there exists another root for the the whole range of parameter space  shown in the 
Fig.~\ref{fig:sturmf} that is located very near to the event horizon (i.e. within 1--1.5 times Schwarzchild radius), but being a centre it is physically untenable to be a sonic point (a critical point through which a physical 
accretion solution, connecting the event horizon with to infinity, 
can pass) and hence has always been justifiably ignored in the literature.
 
\begin{figure}[htb]
	\centering
		\includegraphics[width=0.50\textwidth]{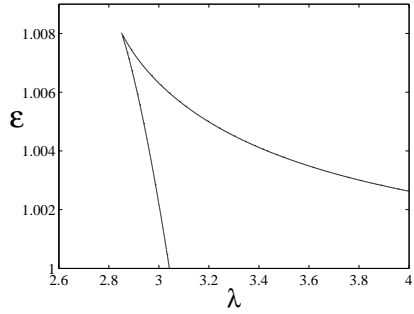}
	\caption{Boundary of transition: Contour line $\det{(S)}=0$ (for $\gamma = 4/3$).}
	\label{fig:sing1}
\end{figure}
The transition boundaries from $n_1$ number of roots to $n_2$ number of roots, in the parameter space, can be more easily obtained using catastrophe theory.
The boundaries of the region in the parameter space permitting transition of number of critical points in this case are associated with saddle-centre bifurcation or merging of a pair of roots of the equation (Eq.\ref{eq43}). Now all these equations are polynomial equations. 
As a general rule the discriminant of a polynomial,
\begin{equation}
	P_n(x)=a_nx^n+a_{n-1}x^{n-1}+\cdots+a_1x+a_0,
\end{equation}
 can be expressed as in terms of its roots, $x_i$'s, as
\begin{equation}
	D=a_n^{n-2}\prod_{i<j}{(x_i-x_j)^2}.   
\end{equation}
The discriminant may be expressed as the determinant of a matrix called 
the Sylvester matrix (see, e.g., 
http://mathworld.wolfram.com/PolynomialDiscriminant.html, and references therein),
\begin{equation}
	S=\left[\begin{array}{lllllllll}
\multicolumn{1}{c}{a_n} & \multicolumn{1}{c}{a_{n-1}} & \multicolumn{1}{c}{a_{n-2}} & \multicolumn{1}{c}{\ldots} & \multicolumn{1}{c}{a_1} & \multicolumn{1}{c}{a_0} & \multicolumn{1}{c}{0\ldots} & \multicolumn{1}{c}{\ldots} & \multicolumn{1}{c}{0} \\ 
\multicolumn{1}{c}{0} & \multicolumn{1}{c}{a_n} & \multicolumn{1}{c}{a_{n-1}} & \multicolumn{1}{c}{a_{n-2}} & \multicolumn{1}{c}{\ldots} & \multicolumn{1}{c}{a_1} & \multicolumn{1}{c}{a_0} & \multicolumn{1}{c}{0\ldots} & \multicolumn{1}{c}{0} \\ 
\multicolumn{1}{c}{\vdots} & \multicolumn{1}{c}{} & \multicolumn{1}{c}{} & \multicolumn{1}{c}{} & \multicolumn{1}{c}{} & \multicolumn{1}{c}{} & \multicolumn{1}{c}{} & \multicolumn{1}{c}{} & \multicolumn{1}{c}{\vdots} \\ 
\multicolumn{1}{c}{0} & \multicolumn{1}{c}{\ldots} & \multicolumn{1}{c}{0} & \multicolumn{1}{c}{a_n} & \multicolumn{1}{c}{a_{n-1}} & \multicolumn{1}{c}{a_{n-2}} & \multicolumn{1}{c}{\ldots} & \multicolumn{1}{c}{a_1} & \multicolumn{1}{c}{a_0} \\ 
\multicolumn{1}{c}{na_n} & \multicolumn{1}{c}{(n-1)a_{n-1}} & \multicolumn{1}{c}{(n-2)a_{n-2}} & \multicolumn{1}{c}{\ldots} & \multicolumn{1}{c}{1a_1} & \multicolumn{1}{c}{0} & \multicolumn{1}{c}{\ldots} & \multicolumn{1}{c}{\ldots} & \multicolumn{1}{c}{0} \\ 
\multicolumn{1}{c}{0} & \multicolumn{1}{c}{na_n} & \multicolumn{1}{c}{(n-1)a_{n-1}} & \multicolumn{1}{c}{(n-2)a_{n-2}} & \multicolumn{1}{c}{\ldots} & \multicolumn{1}{c}{1a_1} & \multicolumn{1}{c}{0} & \multicolumn{1}{c}{\ldots} & \multicolumn{1}{c}{0} \\ 
\multicolumn{1}{c}{\vdots} & \multicolumn{1}{c}{} & \multicolumn{1}{c}{} & \multicolumn{1}{c}{} & \multicolumn{1}{c}{} & \multicolumn{1}{c}{} & \multicolumn{1}{c}{} & \multicolumn{1}{c}{} & \multicolumn{1}{c}{\vdots} \\ 
\multicolumn{1}{c}{0} & \multicolumn{1}{c}{0} & \multicolumn{1}{c}{\ldots} & \multicolumn{1}{c}{0} & \multicolumn{1}{c}{na_n} & \multicolumn{1}{c}{(n-1)a_{n-1}} & \multicolumn{1}{c}{(n-2)a_{n-2}} & \multicolumn{1}{c}{\ldots} & \multicolumn{1}{c}{1a_1} \\ 
\end{array}\right],
\end{equation}
up to a factor.

Putting $n=8$, 
$\det{(S)}$ will be zero on the above mentioned boundaries and actually it is so. 
Here the plot of $\det{(S)}=0$ for the polytropic flow  (i.e. for 
the polynomial in $r_c$ in Eq.\ref{eq43}) in $\cal{E}$--$\lambda$ space is shown in Fig.\ref{fig:sing1}. 
The curve exactly conforms with the corresponding boundary curve  in Fig.\ref{fig:sturmf}, 
drawn on the basis of the previous method. So this procedure may be thought of as a 
much easier alternative to find the multi-critical parametric values; though this method cannot give the exact number 
of critical points in each region of the parameter space.

\section{Discussion}
\label{section6}
\noindent
Our methodology is based on the algebraic form of the first integral 
obtained by solving the radial momentum equation (the Euler equation
to be more specific, since we are confined to the inviscid flow only). 
The structure for such a first integral has to be a formal polynomial with 
appropriate constant co-efficients. 
For general relativistic accretion in the Kerr metric, 
the  expression for the energy first integral can not 
be reduced to such a polynomial form 
(see, e.g., \cite{das-czerny} for the detail form of 
such algebraic expression). Hence, the Sturm's generalized 
chain can not be constructed for such accretion flow. Alternative 
methodology are required to investigate the multi-critical behaviour 
for such kind of accretion. 

Using the method illustrated in this work, it is possible to 
find out how many critical points a transonic black hole accretion 
flow can have. It is thus possible to predict whether such accretion 
flow can have multi-critical properties for a certain specific 
value/domain of the initial boundary conditions. It is, however, 
not possible to investigate, using the eigenvalue analysis as 
illustrated in ~\cite{crd06,gkrd07}, the nature of such critical 
points - i.e., whether they are of saddle type or are of centre type,
since such prediction requires the exact location of the critical 
points (the value of the roots of the polynomial). However, 
the theory of dynamical systems ensures that no two consecutive 
critical points be of same nature (both saddle or both centre). On the 
other hand, our experience predicts (it is rather a documented fact)
that for all kind of multi-critical
black hole accretion, irrespective of the equation of state, the
space time geometry or the flow configuration used, one has two 
saddle type critical points and one centre type critical point
flanked by them (see, e.g., \cite{das-czerny} and \cite{swagata}
for further detail).
Hence if the application of the Sturm's generalized chain ensures the 
presence of three critical points, we can say that out of those three 
critical points, accretion flow will have two saddle type critical 
points, hence a specific subset of the solution having three
roots corresponding to the first integral polynomial, can 
make transonic transition for more than one times, if appropriate 
conditions for connecting the flow through the outer critical 
point and for flow through the inner critical points are available,
see, e.g., \cite{das-czerny} for further discussion. 

In this work we have considered only inviscid accretion. Our methodology
of investigating the multi-critical properties, however, is expected to be 
equally valid for the viscous accretion disc as well. For the viscous flow,
the radial momentum conservation equation involving the first order space 
derivative of the dynamical flow velocity will certainly provide a first 
integral of motion upon integration. Because of the fact that a 
viscous accretion disc is not a non-dissipative system, such constant of
motion, however, can never be identified with the specific energy of the flow.
The integral solution of the radial momentum equation would then be 
an algebraic expression of various flow variables and would perhaps 
involve certain initial boundary conditions as well. Such an algebraic
expression would actually be a constant of motion. What exactly would that 
expression physically signify, would definitely be hard to realize. 
However, one may perhaps arbitrarily parameterize that conserved algebraic 
expression using some astrophysically relevant outer boundary 
conditions, and if such algebraic expressions can finally be reduced,
using the appropriate critical point conditions, to an algebraic 
polynomial form of the critical points, construction of a generalized 
Sturm chain can be made possible to find out how many critical points 
such an accretion flow can have subjected to the specific initial 
boundary condition. Since for accretion onto astrophysical 
black holes, having multiple critical points is a necessary 
(but not sufficient) condition to undergo shock transition, 
one can thus analytically predict, at least to some extent, which 
particular class of viscous accretion disc are susceptible for 
shock formation phenomena. 

Our work, as we believe, can have a broader perspective as well, in the
field of the study of dynamical systems in general. For a first order
autonomous dynamical system, provided one can evaluate
the critical point conditions, the corresponding generalized $n$th
degree algebraic equation involving the position co-ordinate and
one (or more) first integral of motion can be constructed. If such algebraic
equation can finally be reduced to a $n$th degree polynomial with well
defined domain for the constant co efficient, one can easily find out
the maximal number of fixed points of such dynamical systems.
\section*{Acknowledgments}
\noindent
This research has made use of NASA's Astrophysics Data System as well 
as various online encyclopedia. 
SA and SN would like to 
acknowledge the kind hospitality provided by HRI and
by astrophysics project under the XI th plan at HRI., Allahabad, India. 
The work of TKD is partially supported by  
the grant NN 203 380136 provided by the Polish 
academy of sciences and by astrophysics project under the XI th plan at HRI.
RD acknowledges useful discussions with 
S. Ramanna.

\section{Appendix - I : Proof of the Sylvester's theorem:} 
First note that the Sturm sequence
$({f_0},...{f_k})$
is (up to signs) equal to the sequence obtained from the Euclidean
algorithm.  Define a new sequence $({g_0},...,{g_k})$ by
${g_i}={f_i}/{f_k}$ for $i \in \{ 0,...,k \}$. Note that the number
of sign changes in $({f_0}(x),{f_1}(x))$ $($resp.
$(f_{i-1}(x),f_i(x),f_{i+1}(x)))$ and the number of sign changes in
$({g_0}(x),{g_1}(x))$ $($resp.
$({g_{i-1}}(x),{g_i}(x),{g_{i+1}}(x)))$ coincide for any $x$ which
is not a root of $f$. Note also  that the roots of ${g_0}$ are
exactly the roots of $f$ which are not roots of $g$. Observe that
for $i\in{{0,...,k}}$,${g_{i-1}}$ and ${g_i}$ are relatively prime.
We consider, now, how $v(f,g;x)$ behaves when $x$ passes through a
root $c$ of a polynomial ${g_i}$. If $c$ is a root of ${g_0}$,  then
it is not a root of ${g_1}$. We write $f'(c_)>0$ ($resp. <0$) if
$f'$ is positive $($ resp. negative $)$ immediately to the left of
$c$. The sign of $f'(c_{+})$ is defined similarly. Now we recall the following
result: if $R$ is a real closed field, $f\in R[X], a,b\in R$ with $a<b$ and if
the derivative $f'$ is positive (resp. negative) on $]a,b[$, then
$f$ is strictly increasing (resp. strictly decreasing) on $[a,b].$
Then, according to the
signs of $g(c), f'(c_{-})$ and $f'(c_+)$ we have the following  8 cases:

\begin{center}
 {$g(c)>0, f'(c_-)>0, f'(c_+)>0$
\vskip 0.5truecm
\begin{tabular}{|l|l|l|l|r|r|} \hline
{} & {$c_{-}$} & {$c$} & {$c_{+}$} \\ \hline
{$f$}
& {$-$}
&{$0$}
&{$+$}\\ \hline
{$f^{\prime}g$} &
{$+$}
&{}
&{$+$}\\ \hline
\end{tabular}}
\end{center}

\vskip 0.5truecm

\begin{center}
 {$g(c)<0, f'(c_-)>0, f'(c_+)>0$
\vskip 0.5truecm
\begin{tabular}{|l|l|l|l|r|r|} \hline
{} & {$c_{-}$} & {$c$} & {$c_{+}$} \\ \hline
{$f$}
& {$-$}
&{$0$}
&{$+$}\\ \hline
{$f^{\prime}g$} &
{$-$}
&{}
&{$-$}\\ \hline
\end{tabular}}
\end{center}

\vskip 0.5truecm

\begin{center}
 {$g(c)>0, f'(c_-)<0, f'(c_+)>0$
\vskip 0.5truecm
\begin{tabular}{|l|l|l|l|r|r|} \hline
{} & {$c_{-}$} & {$c$} & {$c_{+}$} \\ \hline
{$f$}
& {$+$}
&{$0$}
&{$+$}\\ \hline
{$f^{\prime}g$} &
{$-$}
&{}
&{$+$}\\ \hline
\end{tabular}}
\end{center}

\vskip 0.5truecm

\begin{center}
 {$g(c)<0, f'(c_-)<0, f'(c_+)>0$
\vskip 0.5truecm
\begin{tabular}{|l|l|l|l|r|r|} \hline
{} & {$c_{-}$} & {$c$} & {$c_{+}$} \\ \hline
{$f$}
& {$+$}
&{$0$}
&{$+$}\\ \hline
{$f^{\prime}g$} &
{$+$}
&{}
&{$-$}\\ \hline
\end{tabular}}
\end{center}

\vskip 0.5truecm

\begin{center}
 {$g(c)>0, f'(c_-)>0, f'(c_+)<0$
\vskip 0.5truecm
\begin{tabular}{|l|l|l|l|r|r|} \hline
{} & {$c_{-}$} & {$c$} & {$c_{+}$} \\ \hline
{$f$}
& {$-$}
&{$0$}
&{$-$}\\ \hline
{$f^{\prime}g$} &
{$+$}
&{}
&{$-$}\\ \hline
\end{tabular}}
\end{center}

\vskip 0.5truecm

\begin{center}
 {$g(c)<0, f'(c_-)>0, f'(c_+)<0$
\vskip 0.5truecm
\begin{tabular}{|l|l|l|l|r|r|} \hline
{} & {$c_{-}$} & {$c$} & {$c_{+}$} \\ \hline
{$f$}
& {$-$}
&{$0$}
&{$-$}\\ \hline
{$f^{\prime}g$} &
{$-$}
&{}
&{$+$}\\ \hline
\end{tabular}}
\end{center}

\vskip 0.5truecm

\begin{center}
 {$g(c)>0, f'(c_-)<0, f'(c_+)<0$
\vskip 0.5truecm
\begin{tabular}{|l|l|l|l|r|r|} \hline
{} & {$c_{-}$} & {$c$} & {$c_{+}$} \\ \hline
{$f$}
& {$+$}
&{$0$}
&{$-$}\\ \hline
{$f^{\prime}g$} &
{$-$}
&{}
&{$-$}\\ \hline
\end{tabular}}
\end{center}

\vskip 0.5truecm

\begin{center}
 {$g(c)<0, f'(c_-)<0, f'(c_+)<0$
\vskip 0.5truecm
\begin{tabular}{|l|l|l|l|r|r|} \hline
{} & {$c_{-}$} & {$c$} & {$c_{+}$} \\ \hline
{$f$}
& {$+$}
&{$0$}
&{$-$}\\ \hline
{$f^{\prime}g$} &
{$+$}
&{}
&{$+$}\\ \hline
\end{tabular}}
\end{center}
\vskip 0.5truecm

In every as $x$ passes through $c$, the number of sign changes in
$(f_0(x),f_1(x))$ decreases by $1$ if $g(c)>0$, and increases by $1$
if $g(c)<0$. If $c$ is a root of $g_i$ with $i=1,...k$, then it
is neither a root of $g_{i-1}$ nor a root of $g_{i+1}$, and
$g_{i-1}(c)g_{i+1}(c)<0$, by the definition of the sequence. Passing
through $c$ does not lead to any modification of the number of sign
changes in $(f_{i-1}(x),f_i(x),f_{i+1}(x))$ in this case.

{\bf Proof of the Sturm's theorem:} Using $g=1$ in previous theorem.
\section{Appendix - II: Explicit expressions for the co-efficients for the Sturm chain constructed for the
relativistic axisymmetric accretion}
\begin{eqnarray*}
c_6 &=& \frac{7 a_7^2}{64 a_8}   - a_6/4,\\
c_5 &=& \frac{3 a_6 a_7}{32 a_8}  - \frac{3}{8} a_5\\
c_4 &=& \frac{5 a_5 a_7}{64 a_8}  - a_4 / 2 \\
c_3 &=& \frac{2 a_4 a_7}{32 a_8}  - \frac{5}{8} a_3\\
c_2 &=& \frac{3 a_3 a_7}{64 a_8}  - \frac{3}{4} a_2\\
c_1 &=& \frac{a_2 a_7}{32 a_8} - \frac{7}{8} a_1\\
c_0 &=& \frac{a_1 a_7}{64 a_8}  - a_0
\label{app_2_1}
\end{eqnarray*}
\begin{eqnarray*}
 d_5 &=& \frac{8 a_8 c_4}{c_6} + ( \frac{7 a_7}{c_6} - \frac{8 c_5 a_8}{c_6^2}) c_5 - 6 a_6\\
 d_4 &=& \frac{8 a_8 c_3}{c_6} + ( \frac{7 a_7}{c_6} - \frac{8 c_5 a_8}{c_6^2}) c_4 - 5 a_5\\
 d_3 &=& \frac{8 a_8 c_2}{c_6} + ( \frac{7 a_7}{c_6} - \frac{8 c_5 a_8}{c_6^2}) c_3 - 4 a_4\\
 d_2 &=& \frac{8 a_8 c_1}{c_6} + ( \frac{7 a_7}{c_6} - \frac{8 c_5 a_8}{c_6^2}) c_2 - 3 a_3\\
 d_1 &=& \frac{8 a_8 c_0}{c_6} + ( \frac{7 a_7}{c_6} - \frac{8 c_5 a_8}{c_6^2}) c_1 - 2 a_2\\
 d_0 &=& ( \frac{7 a_7}{c_6} - \frac{8 c_5 a_8}{c_6^2}) c_0 - a_1
\label{app_2_2}
\end{eqnarray*}
\begin{eqnarray*}
 e_4 &=& \frac{d_3 c_6}{d_5} + (\frac{c_5}{d_5} - \frac{d_4 c_6}{d_5^2}) d_4 - c_4\\
 e_3 &=& \frac{d_2 c_6}{d_5} + (\frac{c_5}{d_5} - \frac{d_4 c_6}{d_5^2}) d_3 - c_3\\
 e_2 &=& \frac{d_1 c_6}{d_5} + (\frac{c_5}{d_5} - \frac{d_4 c_6}{d_5^2}) d_2 - c_2\\
 e_1 &=& \frac{d_0 c_6}{d_5} + (\frac{c_5}{d_5} - \frac{d_4 c_6}{d_5^2}) d_1 - c_1\\
 e_0 &=& (\frac{c_5}{d_5} - \frac{d_4 c_6}{d_5^2}) d_0 - c_0
\end{eqnarray*}
\begin{eqnarray*}
f_3 &=& \frac{e_2 d_5}{e_4} + (\frac{d_4}{e_4} - \frac{e_3 d_5}{e_4^2} ) e_3 - d_3\\
f_2 &=& \frac{e_1 d_5}{e_4} + (\frac{d_4}{e_4} - \frac{e_3 d_5}{e_4^2} ) e_2 - d_2\\
f_1 &=& \frac{e_0 d_5}{e_4} + (\frac{d_4}{e_4} - \frac{e_3 d_5}{e_4^2} ) e_1 - d_1\\
f_0 &=&  (\frac{d_4}{e_4} - \frac{e_3 d_5}{e_4^2} ) e_0 - d_0.
\end{eqnarray*}
\begin{eqnarray*}
g_2 &=& \frac{f_1 e_4}{f_3} + (\frac{e_3}{f_3} - \frac{f_2 e_4}{f_3^2}) f_2 - e_2\\
g_1 &=& \frac{f_0 e_4}{f_3} + (\frac{e_3}{f_3} - \frac{f_2 e_4}{f_3^2}) f_1 - e_1\\
g_0 &=& (\frac{e_3}{f_3} - \frac{f_2 e_4}{f_3^2}) f_0 - e_0
\end{eqnarray*}
\begin{eqnarray*}
h_1 &=& \frac{g_0 f_3}{g_2} + ( \frac{f_2}{g_2} - \frac{g_1 f_3}{g_2^2} ) g_1 -f_1\\
h_0 &=& ( \frac{f_2}{g_2} - \frac{g_1 f_3}{g_2^2} ) g_0 - f_0
\end{eqnarray*}
\begin{eqnarray*}
i_0 = (\frac{g_1}{h_1} - \frac{h_0 g_2}{h_1^2} )h_0 - g_0
\end{eqnarray*}

\begin{thebibliography}{10}
\providecommand{\url}[1]{{#1}}
\providecommand{\urlprefix}{URL }
\expandafter\ifx\csname urlstyle\endcsname\relax
  \providecommand{\doi}[1]{DOI \discretionary{}{}{}#1}\else
  \providecommand{\doi}{DOI \discretionary{}{}{}\begingroup
  \urlstyle{rm}\Url}\fi

\bibitem{lt80}
E.P.T. Liang, K.A. Thomson, ApJ. \textbf{240}, 271 (1980)

\bibitem{rb02}
A.K. Ray, J.K. Bhattacharjee, Phys. Rev. E \textbf{66}, 066303 (2002)

\bibitem{ap03}
N.~Afshordi, B.~Paczy\'nski, ApJ. \textbf{592}, 354 (2003)

\bibitem{ray03a}
A.K. Ray, MNRAS \textbf{344}, 83 (2003)

\bibitem{ray03b}
A.K. Ray, MNRAS \textbf{344}, 1085 (2003)

\bibitem{rbcqg05a}
A.K. Ray, J.K. Bhattacharjee.
\newblock A dynamical systems approach to a thin accretion disc and its
  time-dependent behaviour on large length scales.
\newblock eprint {arXiv:astro-ph/0511018v1} (2005)

\bibitem{rbcqg05b}
A.K. Ray, J.K. Bhattacharjee, The Astrophysical Journal \textbf{627}, 368
  (2005)

\bibitem{crd06}
S.~Chaudhury, A.K. Ray, T.K. Das, MNRAS \textbf{373}, 146 (2006)

\bibitem{rbcqg06}
A.K. Ray, J.K. Bhattacharjee, Indian Journal of Physics \textbf{80}, 1123
  (2006).
\newblock Eprint arXiv:astro-ph/0703301

\bibitem{rbcqg07a}
A.K. Ray, J.K. Bhattacharjee, Classical and Quantum Gravity \textbf{24}, 1479
  (2007)

\bibitem{br07}
J.K. Bhattacharjee, A.K. Ray, ApJ. \textbf{668}, 409 (2007)

\bibitem{gkrd07}
S.~Goswami, S.N. Khan, A.K. Ray, T.K. Das, MNRAS \textbf{378}, 1407 (2007)

\bibitem{jkb09}
J.K. Bhattacharjee, A.~Bhattacharya, T.K. Das, A.K. Ray, MNRAS \textbf{398},
  841 (2009).
\newblock Also at arXiv:0812.4793v1 [astro-ph]

\bibitem{az81}
M.A. Abramowicz, W.H. Zurek, ApJ. \textbf{246}, 314 (1981)

\bibitem{boz-pac}
B.~Muchotrzeb, B.~Paczynski, Acta Actron. \textbf{32}, 1 (1982)

\bibitem{boz1}
B.~Muchotrzeb, Acta Astron. \textbf{33}, 79 (1983)

\bibitem{fuk83}
J.~Fukue, PASJ \textbf{35}, 355 (1983)

\bibitem{fuk87}
J.~Fukue, PASJ \textbf{39}, 309 (1987)

\bibitem{fuk04}
J.~Fukue, PASJ \textbf{56}, 681 (2004)

\bibitem{fuk04a}
J.~Fukue, PASJ \textbf{56}, 959 (2004)

\bibitem{lu85}
J.F. Lu, A \& A \textbf{148}, 176 (1985)

\bibitem{lu86}
J.F. Lu, Gen. Rel. Grav. \textbf{18}, 45L (1986)

\bibitem{bmc86}
B.~Muchotrzeb-Czerny, Acta Astronomica \textbf{36}, 1 (1986)

\bibitem{ak89}
M.A. Abramowicz, S.~Kato, ApJ. \textbf{336}, 304 (1989)

\bibitem{abram-chak}
M.A. Abramowicz, S.K. Chakrabarti, ApJ. \textbf{350}, 281 (1990)

\bibitem{ky94}
M.~Kafatos, R.X. Yang, MNRAS \textbf{268}, 925 (1994)

\bibitem{yk95}
R.X. Yang, M.~Kafatos, A \& A \textbf{295}, 238 (1995)

\bibitem{caditz-tsuruta}
D.M. Caditz, S.~Tsuruta, ApJ. \textbf{501}, 242 (1998)

\bibitem{das02}
T.K. Das, ApJ. \textbf{577}, 880 (2002)

\bibitem{bdw04}
P.~Barai, T.K. Das, P.J. Wiita, ApJ. \textbf{613}, 167, L49 (2004)

\bibitem{abd06}
H.~Abraham, N.~Bili\'{c}, T.K. Das, Classical and Quantum Gravity \textbf{23},
  2371 (2006)

\bibitem{dbd07}
T.K. Das, N.~Bili\'c, S.~Dasgupta, JCAP \textbf{06}, 009 (2007)

\bibitem{das-czerny}
T.K. Das, B.~Czerny, New Astronomy \textbf{17}, 254 (2012).

\bibitem{swagata}
S.~Nag, S.~Acharya, A.K. Ray, T.K. Das, New Astronomy \textbf{17}, 285  (2012)

\bibitem{c89}
S.K. Chakrabarti, ApJ. \textbf{347}, 365 (1989)

\bibitem{dpm03}
T.K. Das, J.K. Pendharkar, S.~Mitra, ApJ. \textbf{592}, 1078 (2003)

\bibitem{ila-shu}
A.~Illarionov, R.A. Sunyaev, A \& A \textbf{39}, 205 (1975)

\bibitem{liang-nolan}
E.P.T. Liang, P.L. Nolan, Space. Sci. Rev. \textbf{38}, 353 (1984)

\bibitem{bisikalo}
A.A. Bisikalo, V.M. Boyarchuk, V.M. Chechetkin, O.A. Kuznetsov, D.~Molteni,
  MNRAS \textbf{300}, 39 (1998)

\bibitem{ila}
A.F. Illarionov, Soviet Astron. \textbf{31}, 618 (1988)

\bibitem{ho}
L.C. Ho, in \emph{Observational Evidence For Black Holes in the Universe}, ed.
  by S.K. Chakrabarti (Dordrecht: Kluwer, 1999), p. 153

\bibitem{igu}
I.V. Igumenshchev, M.A. Abramowicz, MNRAS \textbf{303}, 309 (1999)

\bibitem{mkfo84}
R.~Matsumoto, S.~Kato, J.~Fukue, A.T. Okazaki, PASJ \textbf{36}, 71 (1984)

\bibitem{pac87}
B.~Paczy\'{n}ski, Nature \textbf{327}, 303 (1987)

\bibitem{acls88}
M.A. Abramowicz, B.~Czerny, J.P. Lasota, E.~Szuszkiewicz, ApJ. \textbf{332},
  646 (1988)

\bibitem{ct93}
X.~Chen, R.~Taam, ApJ. \textbf{412}, 254 (1993)

\bibitem{abn96}
I.V. Artemova, G.~Bj\"ornsson, I.D. Novikov, ApJ. \textbf{461}, 565 (1996)

\bibitem{nkh97}
R.~Narayan, S.~Kato, F.~Honma, ApJ. \textbf{476}, 49 (1997)

\bibitem{wiita99}
P.J. Wiita, in \emph{Black Holes, Gravitational Radiation and the Universe},
  ed. by B.R. Iyer, B.~Bhawal (Dordrecht: Kluwer, 1999), p. 249

\bibitem{hawley-krolik}
J.F. Hawley, J.H. Krolik, ApJ. \textbf{548}, 348 (2001)

\bibitem{armitage}
P.J. Armitage, C.S. Reynolds, J.~Chiang, ApJ. \textbf{648}, 868 (2001)

\bibitem{alp97}
M.A. Abramowicz, A.~Lanza, M.J. Percival, ApJ \textbf{479}, 179 (1997)

\bibitem{man2000}
T.~Manmoto, ApJ \textbf{534}, 734 (2000)

\bibitem{stark}
H.~Stark, \emph{An Introduction to Number Theory}.
\newblock ISBN 0-262-69060-8 (MIT Press, 1978)

\bibitem{RAG}
J.~Bochnak, M.~Coste, M.F. Roy, \emph{Real Algebraic Geometry} (Springer, 1991)

\end{thebibliography}

\end{document}